\documentstyle[aps,preprint,floats,psfig]{revtex}
\begin{document}
\newcommand {\be}{\begin{equation}}
\newcommand {\ee}{\end{equation}}
\newcommand {\bea}{\begin{eqnarray}}
\newcommand {\eea}{\end{eqnarray}}
\newcommand {\nn}{\nonumber}

\draft
%
%
%
%

\title{Temperature dependence of impurity bound states in
$\rm d_{x^2 - y^2}$-wave superconductors
}

\author{Stephan Haas and Kazumi Maki}
\address{Department of Physics and Astronomy, University of Southern
California, Los Angeles, CA 90089-0484
}

\date{\today}
\maketitle

\begin{abstract}
We study the evolution with temperature
of quasiparticle bound states around non-magnetic impurities 
in $\rm d_{x^2 - y^2}$-wave superconductors. The
associated local density of states has a fourfold symmetry 
which has recently been observed in Zn-doped Bi2212
using scanning tunneling microscopy (STM).  
From the corresponding Bogoliubov-de Gennes equation we find
that with increasing temperature the 
magnitude of the bound state energy increases and
the amplitude of the fourfold contribution to the spinor wave 
functions decreases. In the pseudogap regime above $\rm T_c$ the 
fourfold angular dependence of the local tunneling conductance 
persists as long as the superconducting fluctuations 
are sufficiently strong to support a finite local order 
parameter. Once the gap function vanishes completely, the
angular structure of the bound state
wave function becomes featureless. These effects should be observable
in STM studies of impurity doped high-temperature superconductors. 
  
\end{abstract}
\pacs{}

{\it Introduction:}
It is well known that impurity doping, such as Zn or Ni substitution
in the high-temperature superconductors LSCO, YBCO, and BSCCO, is
a useful tool in demonstrating the underlying nodal structure 
of their order parameter. It  also
provides a semi-quantitative test of the BCS theory for 
$\rm d_{x^2 - y^2}$-wave superconductors.\cite{hirschfeld,haas,sun,maki}
In particular, an examination of the thermodynamic and transport properties
in these compounds suggests that Zn impurities can be modeled with a 
scattering potential in the unitary limit.\cite{sun,maki}
On the other hand, little is known experimentally about the local 
structure of the impurity bound states around the Zn sites despite
numerous theoretical studies on this question.\cite{salkola,haas2}
Recently, Pan {\it et al.} have provided the first scanning tunneling
microscope (STM) images of the local tunneling conductance around
Zn impurities in Bi2212 at low temperatures, T $\approx$ 4.3 K, and fixed at 
$\pm E$, where $E \approx \Delta/30$ is the binding energy of
the impurity bound state.\cite{pan}
The corresponding wave function was shown to exhibit a fourfold angular 
symmetry associated with the underlying $\rm d_{x^2 - y^2}$-wave order
parameter.
We have found that 
this quasiparticle bound state wave
function around impurity sites can be described within the formalism
of the Bogoliubov
- de Gennes (BdG) equations for $\rm d_{x^2 - y^2}$-wave superconductors.
\cite{haas2} Within this framework, the main features of these observations,
in particular the characteristic angular patterns seen in
the STM images of Zn-doped
BSCCO, were reproduced. 

While this previous work was based on an analysis of the BdG equations
at zero temperature, the 
thermal evolution of the properties of the impurity bound state 
is of particular interest. 
Here we will study 
how the angular STM patterns around impurities may provide valuable
information regarding the nature of the pseudogap in the 
temperature regime above $T_c$. 
We will assume that
the origin of the pseudogap is dominated by standard 
superconducting fluctuations.\cite{won,varlamov,martin,morr} 
This assumption is motivated
by recent angle-resolved 
photoemission spectroscopy (ARPES) measurements
on underdoped BSCCO which have clearly indicated
an angular dependence of the pseudogap feature proportional to 
$\cos^2{(2 \phi)}$, analogous to $|\Delta({\bf k})|^2$ in the 
$\rm d_{x^2 - y^2}$-wave superconducting phase of this compound.
\cite{ding}  
In the following, we will address the consequences 
of the pseudogap phenomenon on the bound state wave function 
in a semi-phenomenological manner. In particular, we will investigate
how  the characteristic
fourfold symmetry pattern in the wave function gradually disappears
as the temperature is increased. 

{\it Bogoliubov - de Gennes equations:}
Let us start our analysis by 
examining the BdG equations for the spinor wave functions in
a $\rm d_{x^2 - y^2}$-wave superconductor, given by\cite{haas2}  
\bea
E u({\bf r}) &=& \left( -\frac{\nabla^2}{2 m} - \mu - V ({\bf r}) \right)
u({\bf r}) + \frac{1}{p_F^2} \Delta (\partial_x^2 - \partial_y^2 )
v({\bf r}),\\
E v({\bf r}) &=& - \left( -\frac{\nabla^2}{2 m} - \mu - V ({\bf r}) \right)
v({\bf r}) + \frac{1}{p_F^2} \Delta (\partial_x^2 - \partial_y^2 )
u({\bf r}),
\eea
where $\mu$ is the chemical potential, $p_F$ is the Fermi momentum, and 
$V ({\bf r}) = V_0 \delta^2 ({\bf r}) > 0$ is an isotropic
impurity scattering potential, centered at ${\bf r} = 0$.
In previous work it was shown that the spinor wave functions $u({\bf r})$
and $v({\bf r})$ can be expanded in terms of Bessel functions of the 
first kind, leading to the Ansatz\cite{haas2}
\bea
u({\bf r}) &=&  A \exp{(- \gamma r)} \left( J_0(p_F r) + \sqrt{2} \beta
J_4(p_F r) \cos{(4 \phi)} \right),\\
 v({\bf r}) &=&  \sqrt{2} A \alpha \exp{(- \gamma r)} J_2(p_F r) \cos{(2 \phi)},
\eea
where $J_l(p_F r)$ are Bessel functions of the first kind,
$\alpha$, $\beta$, and $\gamma$ are variational 
parameters, and $A$ is a global normalization factor.
In this expansion, only
the leading-order angle-dependent terms have 
been retained. 

Since the spinor wave function $v({\bf r})$ does not contain an 
s-wave component, it can be eliminated from Eq. (1), yielding  
\bea
E u({\bf r}) = \left( K - V
+ \frac{\Delta^2 (1 + \cos{(4 \phi)})}{2 ( K + E)} \right)
u({\bf r}),
\eea
where the contributions of the kinetic energy and the impurity
scattering potential are given by
\bea
K &\equiv &  \frac{\int_0^{\infty} dr r \left[
\left( \partial_r \exp{(-\gamma r)}
J_l(p_F r)\right)^2 + \left( l \exp{(-\gamma r)} J_l(p_F r)/r \right)^2
\right]}
{2m \int_0^{\infty} dr r \left( \exp{(-\gamma r)} J_l(p_F r)\right)^2}
-\mu
\simeq \frac{\gamma^2}{2 m},\\
V &\equiv & \frac{\int_0^{\infty} dr r \exp{(-2 \gamma r)} J_0^2(p_F r)
V({\bf r})}
{\int_0^{\infty} dr r \exp{(-2 \gamma r)} J_0^2(p_F r)}
\simeq (2 \pi \gamma p_F) \int_0^{\infty} dr r \exp{(-2 \gamma r)}
J_0^2(p_F r) V ({\bf r}).
\eea
Note that the contribution of the impurity potential 
$V$ does not appear in the 
denominator of the last term in Eq. (5) because
the spinor wave function
$v({\bf r})$ does not have an s-wave component.
Furthermore, the kinetic energy contribution reduces to 
$\gamma^2/2 m$ for all $J_l(p_F r)$ with $l \ll p_F/\gamma$.\cite{haas2} 
The two parameters of interest which will be determined in the following
are the temperature-dependent binding energy $E(T)$ and the 
coefficient of the fourfold symmetric term $\beta (T)$. 
Because $u({\bf r})$ contains
two orthogonal angular components, and the kinetic energy $K$ is 
independent of angular momentum, it follows that 
\bea 
\left( E - K + V - \frac{\Delta^2}{2(E + K)} \right)
- \frac{\beta \Delta^2}{2\sqrt{2}(E + K)} & = & 0, \\
\left( E - K - \frac{\Delta^2}{2(E + K)} \right)\beta
- \frac{\Delta^2}{2\sqrt{2}(E + K)} & = & 0.
\eea
Solving these two equations for $E$ and $\beta$, one obtains
\bea
\beta &=& \frac{\Delta^2}{2\sqrt{2}(E^2 - K^2 - \Delta^2/2)},\\
\Delta^4/8 &=& \left( E^2 - K^2 + V(K + E) - \Delta^2/2 \right)
\left( E^2 - K^2 - \Delta^2/2 \right).
\eea
The quartic equation for $E$ contains 
one root which will determine the bound state energy.

To make further progress, 
we will assume that
(i) the impurity scattering is in the unitary limit, $V_0 \agt \Delta_0$,
leading to a 
small binding energy E especially at low temperatures, 
(ii) the energy contributions
$K$ and $V$ are independent of 
temperature in the regime of interest,
$T \in [0, 3 T_c]$,
and (iii) the local amplitude of the superconducting
gap function $\Delta (T)$ is small but finite in the 
pseudogap regime above $T_c$. 

{\it Superconducting regime:}
Let us first consider the zero-temperature limit, assuming that 
the impurity scattering potential is in the unitary scattering 
limit. In this case, $E \rightarrow 0$, leading to
\bea
V \rightarrow \frac{1}{K^2 + \Delta_0^2/2} \left[ K(K^2 + \Delta_0^2) 
+ \Delta_0^4/8K \right].
\eea
Following the assumption of temperature-independent
energy contributions
$K$ and $V$, we can set $K = a \Delta_0$, where $\Delta_0$
is the gap amplitude at $T = 0$ , and $a$ is a proportionality
constant whose physical range can be inferred from experiments. 
It then follows that
\bea
V \simeq \frac{1 + 8 a^2(1 + a^2)}{4a (1 + 2 a^2)}\Delta_0.
\eea
Furthermore, it is known that at zero temperature
$\beta < 0$, leading to the characteristic fourfold symmetric
patterns in the local density of states around impurity
sites that have been observed in recent
STM measurements.\cite{pan} It is 
therefore natural to postulate that $\beta < 0$ for all temperatures,
imposing the constraint: $a >  2^{-3/4} = 0.5946$.
Therefore, we will only consider the
physical parameter regime $a >  2^{-3/4}$.
For example, we can set $a = 2/3$, yielding
$\beta = -9/(17\sqrt{2}) = -0.3743$ at T=0. We have verified
that the finite-temperature
properties of the bound state
that will be discussed in the following do not depend strongly on
the choice of $a$. 

{\it Normal state:}
In the opposite limit, $T \gg T_c$, the gap function vanishes
$\Delta^2 (T \rightarrow \infty)
\rightarrow 0$. Assuming that $K$ and $V$ are basically 
unaffected by temperature, this gives
\bea
E(T \rightarrow \infty )
 = K - V = - \frac{(1 + 4 a^2)}{4a (1 + 2 a^2)}\Delta_0.
\eea
Hence, the magnitude of the impurity bound state energy $E$
increases with increasing temperature.
The parameter $\beta$ vanishes in this limit, indicating that the 
bound state wave function loses its fourfold angular symmetry 
pattern at high temperatures when $\Delta^2(T) \rightarrow 0$.
 
{\it Pseudogap regime:}
In the pseudogap regime the amplitude of the local order 
parameter, $| \Delta(T)|$, is assumed to be small 
but finite, and by perturbing about the limit $T \rightarrow \infty$
one finds to leading order that
\bea
E(T > T_c) \approx E(T \rightarrow \infty ) + \frac{\Delta^2(T)}{2(K + E)}
\simeq  - \frac{(1 + 4 a^2)}{4a (1 + 2 a^2)}\Delta_0
+ \frac{2a (1 + 2 a^2) \Delta^2(T) }{(8 a^4 - 1) \Delta_0},
\eea
and
\bea
\beta(T > T_c) \approx - \frac{16 a^2 (1 + 2 a^2)^2 \Delta^2(T) }
{2\sqrt{2} (64 a^6(1 + a^2) - (1 + 8a^2))\Delta_0^2 }.
\eea
Hence the magnitude of the binding energy $E(T)$ decreases as
$T \rightarrow 0$ due to the progressive opening of the energy gap. 

Provided that the pseudogap above $T_c$
arises mainly due to superconducting fluctuations
\cite{won,varlamov,martin,morr}, 
it is straightforward to incorporate its physical consequences
by assuming a small but finite amplitude of the local order parameter
above $T_c$.
This can be modeled by
\bea
\Delta^2(T)=\frac{1}{2}\left[\Delta^2_0(1-(T/T_c)^3)^2
+ {\sqrt{\Delta^4_0(1 - (T/T_c)^3)^4 + C }} \right],
\eea
where $C = 0.027572$. The coefficient $C$ is
obtained from 
\bea
\frac{C}{4 \Delta^4_0((T/T_c)^3 - 1)^2} 
\simeq \frac{2\pi T}{m \xi^2_0 \ln{(T/T_c)}},
\eea
where the right hand side is
the spatial average of the fluctuation order parameter, and
$\xi_0 = 73 (3) v^2/2(2\pi T)^2$. The approximate
ratio $T_c/E_F \simeq 0.03$
in Bi2212 has been deduced from the low-temperature behavior
of the thermal conductivity.\cite{chiao} This interpolation 
formula for $\Delta^2(T)$ has already
proven successful in the analysis 
of the excess Dingle temperatures in the vortex state of the 
$\kappa - (ET)_2$ salts.\cite{maki2,ito}

In Fig. 1, the temperature-dependent gap amplitude is plotted 
along with the energy of the bound state and the coefficient 
of the fourfold symmetry term in the spinor wave function 
$u({\bf r})$. The gap amplitude at zero-temperature has been
set equal to unity, $\Delta_0 \equiv 1$. At very small temperatures,
we observe that
$\Delta(T \rightarrow 0 ) \rightarrow 1$, $E (T \rightarrow 0)
\rightarrow 0$, and $\beta (T \rightarrow 0) \rightarrow  
-(\sqrt{2}(a^2 +1))^{-1}$, as expected from the discussion 
of the $T \rightarrow 0$ limit.  In the limit $T \rightarrow \infty$, we
find that $\Delta(T \rightarrow 0 ) \rightarrow 0$,
$E (T \rightarrow -(1+4a^2)(4a + 8a^3)^{-1}$, and
$\beta (T \rightarrow 0) \rightarrow 0$, implying that the 
magnitude of the binding energy increases with temperature, and 
that the fourfold angular features in the spinor wave function 
disappear 
at high temperatures. The intermediate-temperature
regime of $E(T)$ and $\beta (T)$ in Fig. 1 has been
determined by a numerical solution of the coupled equations
(10) and (11). In the fluctuation regime
around $T_c$, the coefficient $\beta (T \approx T_c )$ is found
to be appreciable. Therefore, remnants of the characteristic
$\cos^2{(2 \phi)}$-dependence in the square of the bound state 
spinor wave function $u({\bf r})$ should be reflected 
in the local tunneling density of states, measured by STM
experiments in the pseudogap regime above $T_c$.
In Fig. 2, the corresponding temperature evolution of 
$|u({\bf r})|^2$ is shown.

{\it Conclusions:}
In summary, we have studied the effect of temperature on the 
bound state wave function around impurities in 
$\rm d_{x^2 - y^2}$-wave superconductors. Within the framework 
of the BdG equations,
the magnitude of the binding energy is found to increase with 
temperature, but stays on the order of $\Delta_0$. On the other 
hand, the fourfold contribution to the local density of states
disappears gradually with increasing temperature. Therefore it 
appears to be possible to take a ``snapshot" of the pseudogap 
by STM imaging. The present analysis may also be applicable to other 
unconventional superconductors, such as the layered organic
superconductors $\rm \kappa-(ET)_2 Cu[N(CN)_2]Br$ and
$\rm \kappa-(ET)_2 Cu(NCS)_2$, and $\rm Sr_2RuO_4$.

We thank Alexander Balatsky, Ivar Martin,  Bruce Normand, and 
Hyekyung Won for useful discussions.
S. H. acknowledges the Zumberge foundation for financial support.

\newpage

\begin{figure}[h]
\centerline{\psfig{figure=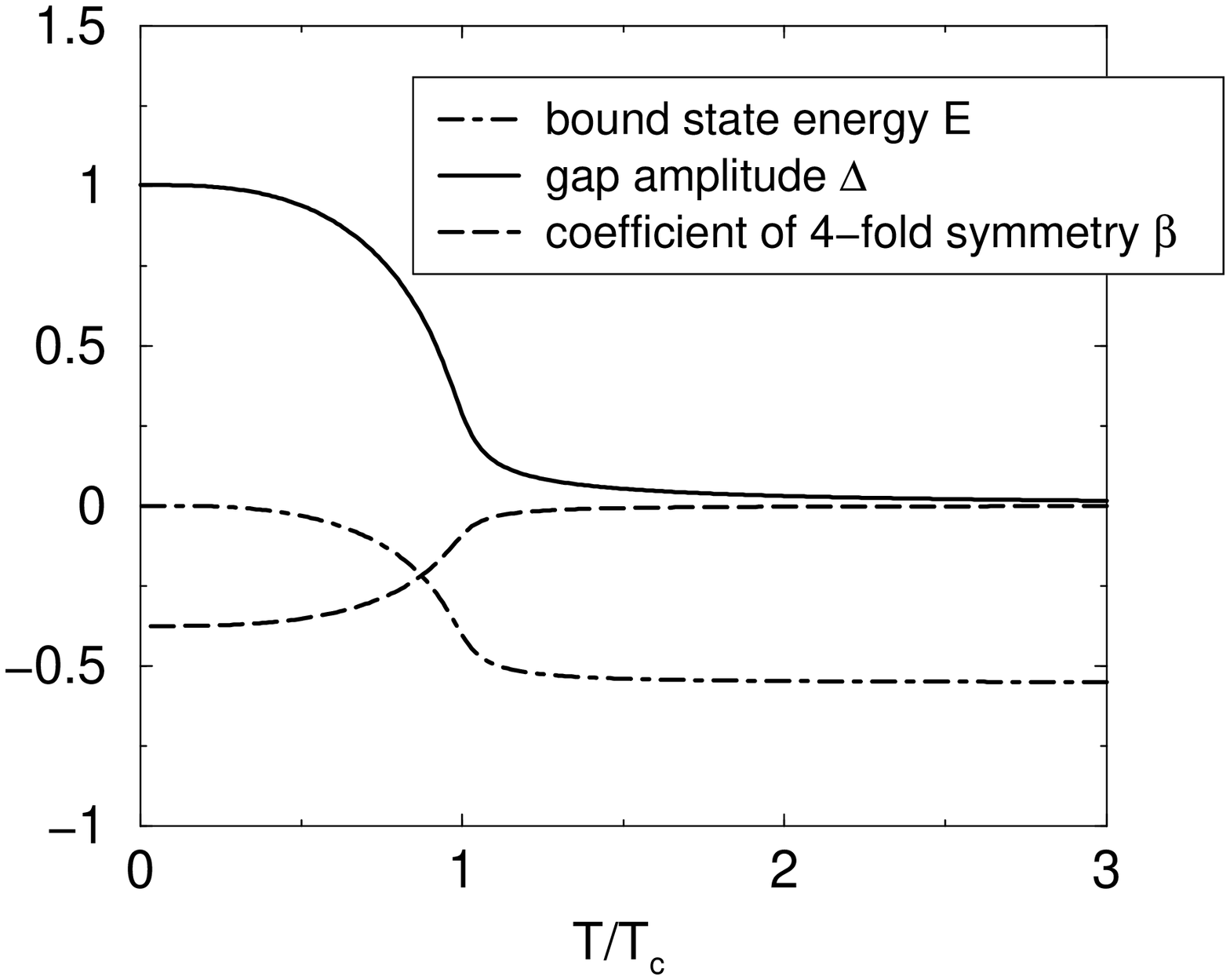,width=9cm,angle=0}}
\vspace{0.3cm}
\caption{
Temperature dependence of the gap amplitude $\Delta$ (solid line),
the bound state energy $E$ (dashed line), and the coefficient of
the fourfold symmetry term in the bound state wave function $\beta$
(dot-dashed line). The proportionality constant $a$ has been set
to $a = 2/3$.
}
\end{figure}
\begin{figure}[h]
\vspace{-0.9cm}
\centerline{\psfig{figure=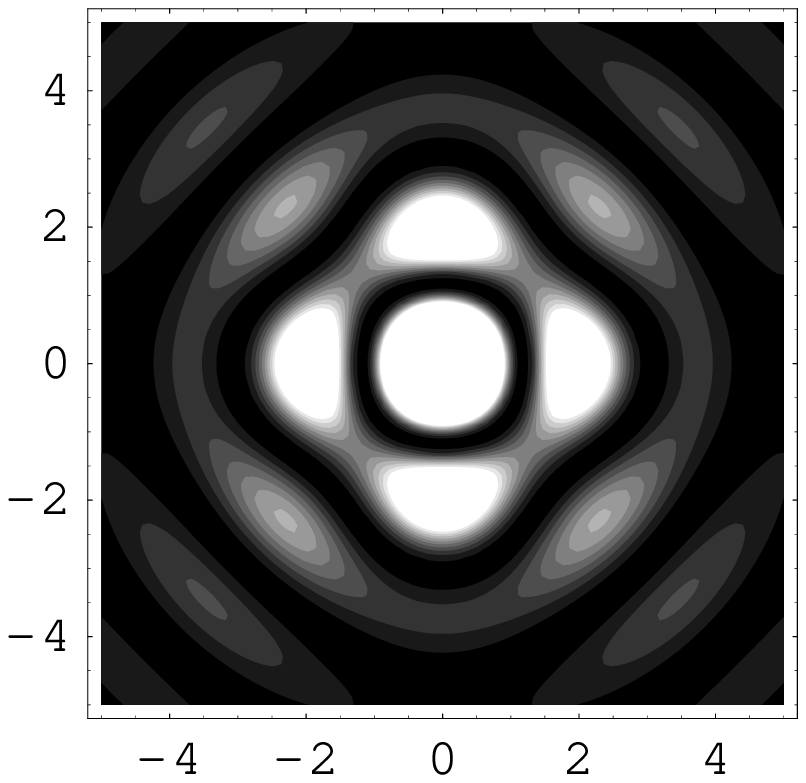,width=6cm,angle=0}
\psfig{figure=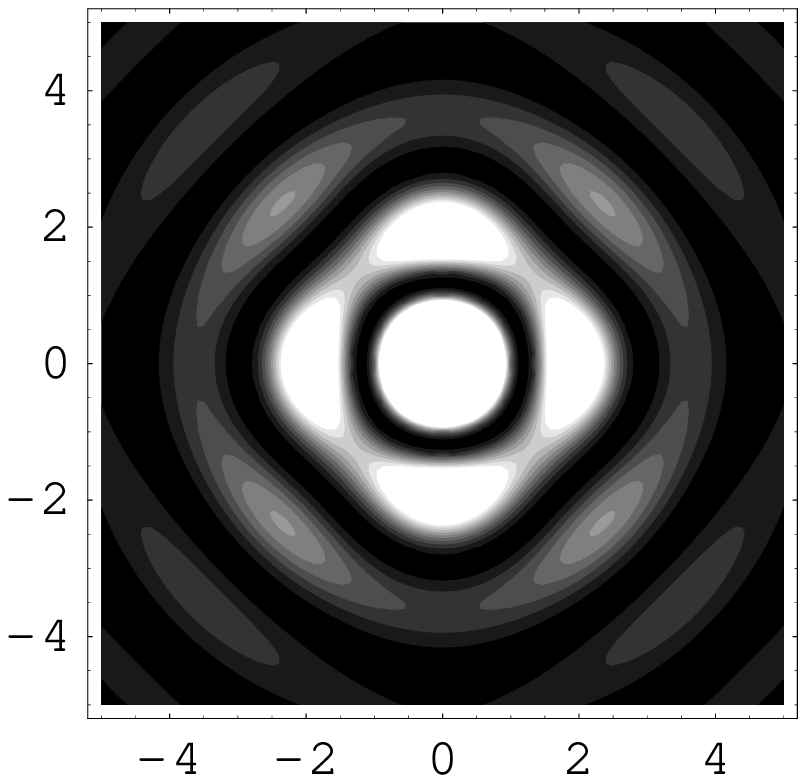,width=6cm,angle=0}
\psfig{figure=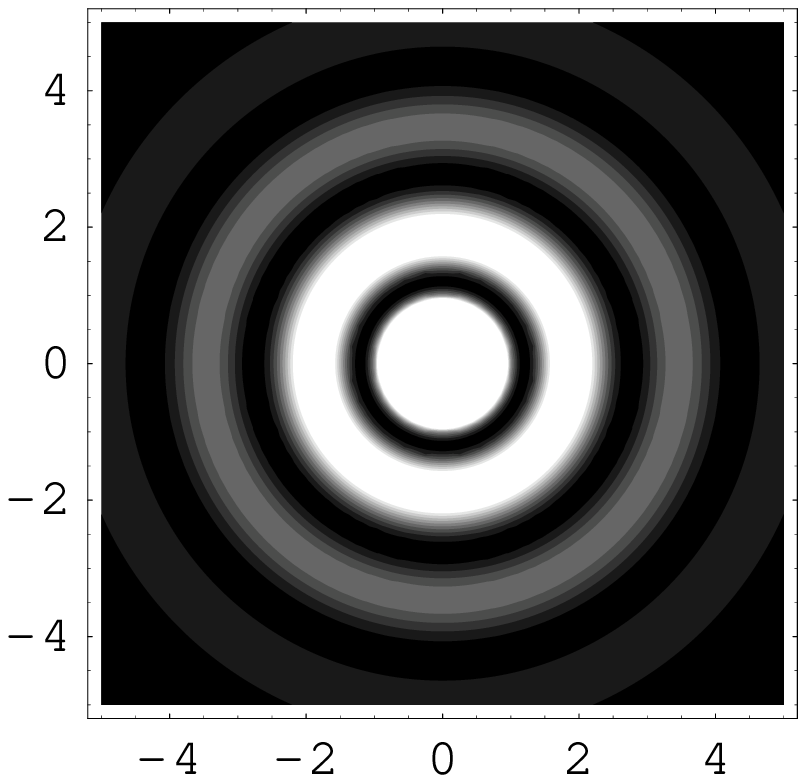,width=6cm,angle=0}}
\vspace{0.3cm}
\caption{
Temperature dependence of the bound state wave function,
$|u({\bf r})|^2$, localized around a
strong-scattering impurity in a $\rm d_{x^2 - y^2}$-wave
superconductor. left: superconducting regime ($T = 0$),
center: pseudogap regime ($T > T_c$), right:
normal state ($T = \infty $).
}
\end{figure}

\end{document}